
\magnification=\magstep1       	
\font\bigbold=cmbx10 scaled 1200   
\font\tenbm=cmmib10		
\newfam\bmfam\textfont\bmfam=\tenbm
\def\bmit{\fam\bmfam\tenbm}
\mathchardef\Theta="7102
\mathchardef\delta="710E

\newcount\EQNO      \EQNO=0
\newcount\FIGNO     \FIGNO=0
\newcount\REFNO     \REFNO=0
\newcount\SECNO     \SECNO=0
\newcount\SUBSECNO  \SUBSECNO=0
\newcount\FOOTNO    \FOOTNO=0
\newbox\FIGBOX      \setbox\FIGBOX=\vbox{}
\newbox\REFBOX      \setbox\REFBOX=\vbox{}
\newbox\RefBoxOne   \setbox\RefBoxOne=\vbox{}

\expandafter\ifx\csname normal\endcsname\relax\def\normal{\null}\fi

\def\Eqno{\global\advance\EQNO by 1 \eqno(\the\EQNO)%
    \gdef\label##1{\xdef##1{\nobreak(\the\EQNO)}}}
\def\Fig#1{\global\advance\FIGNO by 1 Figure~\the\FIGNO%
    \global\setbox\FIGBOX=\vbox{\unvcopy\FIGBOX
      \narrower\smallskip\item{\bf Figure \the\FIGNO~~}#1}}
\def\Ref#1{\global\advance\REFNO by 1 \nobreak[\the\REFNO]%
    \global\setbox\REFBOX=\vbox{\unvcopy\REFBOX\normal
      \smallskip\item{\the\REFNO .~}#1}%
    \gdef\label##1{\xdef##1{\nobreak[\the\REFNO]}}}
\def\Section#1{\SUBSECNO=0\advance\SECNO by 1
    \bigskip\leftline{\bf \the\SECNO .\ #1}\nobreak}
\def\Subsection#1{\advance\SUBSECNO by 1
    \medskip\leftline{\bf \ifcase\SUBSECNO\or
    a\or b\or c\or d\or e\or f\or g\or h\or i\or j\or k\or l\or m\or n\fi
    )\ #1}\nobreak}
\def\Footnote#1{\global\advance\FOOTNO by 1 
    \footnote{\nobreak$\>\!{}^{\the\FOOTNO}\>\!$}{#1}}
\def\SameFootnote{$\>\!{}^{\the\FOOTNO}\>\!$}

\def\References{\bigskip\centerline{\bf REFERENCES}
                \smallskip\copy\REFBOX}
\def\NewRefPage{\setbox\RefBoxOne=\vbox{\unvcopy\REFBOX}%
		\setbox\REFBOX=\vbox{}%
		\def\References{\bigskip\centerline{\bf REFERENCES}
                		\nobreak\smallskip\nobreak\copy\RefBoxOne
				\vfill\eject
				\smallskip\copy\REFBOX}%
		\def\NewRefPage{}}



\nopagenumbers

\def\today{\number\day\space\ifcase\month\or
  January\or February\or March\or April\or May\or June\or
  July\or August\or September\or October\or November\or December\fi
  \space\number\year}
\noindent{gr-qc/9610063 \hfill 30 September 1996}
\bigskip\bigskip

\centerline{\bigbold GRAVITY AND SIGNATURE CHANGE}
\bigskip\bigskip

\centerline{Tevian Dray
\Footnote{Permanent address is Oregon State University.}
}
\centerline{\it Dept.\ of Physics and Mathematical Physics,
		University of Adelaide, Adelaide, SA 5005, AUSTRALIA}
\centerline{\it School of Physics and Chemistry, Lancaster University,
		Lancaster LA1 4YB, UK}
\centerline{\it Department of Mathematics, Oregon State University,
		Corvallis, OR  97331, USA}
\centerline{\tt tevian{\rm @}math.orst.edu}

\medskip
\centerline{George Ellis}
\centerline{\it Department of Applied Mathematics, University of Cape Town,
		Rondebosch 7700, SOUTH AFRICA}
\centerline{\tt ellis{\rm @}maths.uct.ac.za}

\medskip
\centerline{Charles Hellaby
\Footnote{Permanent address is University of Cape Town.}
}
\centerline{\it School of Physics and Chemistry, Lancaster University,
		Lancaster LA1 4YB, UK}
\centerline{\it Department of Applied Mathematics, University of Cape Town,
		Rondebosch 7700, SOUTH AFRICA}
\centerline{\tt cwh{\rm @}maths.uct.ac.za}

\medskip
\centerline{Corinne A. Manogue
$\>\!{}^1\>\!$
}
\centerline{\it Dept.\ of Physics and Mathematical Physics,
		University of Adelaide, Adelaide, SA 5005, AUSTRALIA}
\centerline{\it School of Physics and Chemistry, Lancaster University,
		Lancaster LA1 4YB, UK}
\centerline{\it Department of Physics, Oregon State University,
		Corvallis, OR  97331, USA}
\centerline{\tt corinne{\rm @}physics.orst.edu}

\bigskip\bigskip\bigskip
\centerline{\bf ABSTRACT}
\midinsert
\narrower\narrower\noindent
The use of proper ``time'' to describe classical ``spacetimes'' which contain
both Euclidean and Lorentzian regions permits the introduction of smooth
(generalized) orthonormal frames.  This remarkable fact permits one to
describe both a variational treatment of Einstein's equations and distribution
theory using straightforward generalizations of the standard treatments for
constant signature.
\endinsert

\vfill
\eject

\headline={\hss\rm -~\folio~- \hss}     

\Section{INTRODUCTION}

A {\it signature-changing spacetime} is a manifold which contains both
Euclidean~%
\Footnote{Due to the frequent misuse of the word {\it Riemannian} to describe
manifolds with metrics of any signature, we instead use {\it Euclidean}
to describe manifolds with a positive-definite metric and {\it Lorentzian} for
the usual signature of relativity.  This is not meant to imply flatness in the
former case, nor curvature in the latter.}
and Lor\-entzian regions.  Signature-changing metrics must be either degenerate
(vanishing determinant) or discontinuous, but Einstein's equations implicitly
assume that the metric is nondegenerate and at least continuous.~%
\Footnote{This can be weakened
\Ref{Robert P. Geroch and Jennie Traschen,
{\it Strings and other distributional sources in general relativity},
Phys.\ Rev.\ {\bf D36}, 1017-1031 (1987).}
to allow locally integrable metrics admitting a square-integrable weak
derivative.  Discontinuous metrics do not satisfy this condition.}
Thus, in the presence of signature change, it is not obvious what ``the''
field equations should be.

For discontinuous signature-changing metrics, one can derive such
equations from a suitable variational principle
\Ref{Tevian Dray,
{\it Einstein's Equations in the Presence of Signature Change},
J. Math.\ Phys.\ (in press).}\label\Einstein
.  This turns out to follow from the existence in this case of a natural
generalization of the notion of orthonormal frame.  The standard theory of
tensor distributions, as well as the usual variation of the Einstein-Hilbert
action, can both be expressed in terms of orthonormal frames, and thus
generalize in a straightforward manner to these models.  No such derivation is
known for continuous signature-changing metrics.  Our key point is that
although signature change requires the metric to exhibit some sort of
degeneracy, there is in the discontinuous case a more fundamental field,
namely the (generalized) orthonormal frame, which remains smooth.

We introduce here two simple examples in order to establish our terminology.
A typical {\it continuous} signature-changing metric is
$$ds^2 = t\,dt^2 + a(t)^2\,dx^2 \Eqno$$ \label\Continuous
whereas a typical {\it discontinuous} signature-changing metric is
$$ds^2 = {\rm sgn}(\tau) \, d\tau^2 + a(\tau)^2\,dx^2 \Eqno$$
\label\Discontinuous
Away from the surface of signature change at $\Sigma=\{t=0\}=\{\tau=0\}$,
these metrics are related by a smooth coordinate transformation, with $\tau$
denoting proper ``time'' away from $\Sigma$.  However, since
$d\tau=\sqrt{|t|}\,dt$, the notions of smooth tensors associated with these
coordinates are different at $\Sigma$, corresponding to different
differentiable structures.

We argue here in favor of the discontinuous metric approach, both physically
and mathematically.  Physically, because of the fundamental role played by
proper time.  Mathematically, because of the geometric invariance of the unit
normal to the surface of signature change.  The resulting (generalized)
orthonormal frames provide a clear path leading to a straightforward
generalization of both Einstein's equations and the theory of tensor
distributions.

\Section{PHYSICS}

A standard tool in the description of physical processes is the introduction
of an orthonormal frame.  Physical quantities can be expressed in terms of
tensor components in an orthonormal frame, corresponding to measurements using
proper distance and proper time.

For example, when studying a scalar field on signature-changing backgrounds
such as \Continuous\ or \Discontinuous, it is important to know the value of
the canonical momentum at the boundary, which is essentially the derivative of
the field with respect to proper time.  Furthermore, the well-posedness of the
initial-value problem in the Lorentzian region tells us that the canonical
momentum will be well-behaved at $\Sigma$ if it is well-behaved at early
times.

Continuous signature-changing metrics necessarily have vanishing determinant
at the surface of signature change, which prevents one from defining an
orthonormal frame there.  The situation is different for signature-changing
metrics such that proper ``time'' $\tau$ is an admissible coordinate.
Although the metric is necessarily discontinuous, 1-sided orthonormal frames
can be smoothly joined at $\Sigma$.  Remarkably, the resulting {\it
generalized orthonormal frame} is smooth, and is as orthonormal as possible.
In fact, requiring not only that the 1-sided induced metrics on $\Sigma$, but
also the 1-sided orthonormal frames, should agree at $\Sigma$ implies that
either the full metric is continuous (and nondegenerate) or that the signature
changes.

Such frames can be used to derive Einstein's field equations from the
Einstein-Hilbert action, obtained for constant signature by integrating the
Lagrangian density
$${\cal L} = g_{ac} \, R^c{}_b \wedge {*} \! \left( e^a \wedge e^b \right)
  \Eqno$$
where
$$R^a{}_b = d\omega^a{}_b + \omega^a{}_c \wedge \omega^c{}_b \Eqno$$
are the curvature 2-forms and $*$ denotes the Hodge dual.  Varying this action
with respect to the metric-compatible connection $\omega$ leads to the further
condition that $\omega$ be torsion-free, while varying with respect to the
(arbitrary) frame $e$ leads to Einstein's equations.  In the presence of a
boundary, one obtains~%
\Footnote{A surface term (the trace of the extrinsic curvature) must be added
to the Einstein-Hilbert action in the presence of boundaries; this has nothing
to do with signature change.}
(in vacuum) the Darmois junction condition
\Ref{G Darmois,
{\bf M\'emorial des Sciences Math\'ematiques},
Fascicule 25, Gauthier-Villars, Paris, 1927.}%
, ~~  namely that the extrinsic curvature of the boundary must be the same as
seen from each side.  (In general, one obtains the usual Lanczos equation
\Ref{C. Lanczos, Phys.\ Z. {\bf 23}, 539 (1922);
\hfill\break
C. Lanczos, Ann.\ Phys.\ (Leipzig) {\bf 74}, 518 (1924).}
relating the stress tensor of the boundary to the discontinuity in the
extrinsic curvature.)

The above derivation of Einstein's equations requires that the connection
1-forms admit (1-sided) limits to the boundary.  For continuous
signature-changing metrics, the connection 1-forms typically blow up at the
boundary, but for discontinuous signature-changing metrics in a (1-sided)
orthonormal frame they don't.~%
\Footnote{This will be the case if each 1-sided manifold-with-boundary has a
well-defined connection, as for instance when glueing manifolds together or,
on the Lorentzian side, when starting from well-posed initial data.}
It thus seems reasonable to propose that ``Einstein's equations'' for
signature-changing manifolds should be obtained by varying the (piecewise
extension of) the above action
\Footnote{There are a number of relative sign ambiguities between regions of
different signature, so that the relative sign in the action --- and hence in
the boundary conditions --- can be chosen arbitrarily.}
with respect to the (generalized) orthonormal frame.  As expected, one obtains
Einstein's equations separately in the two regions together with the Darmois
junction conditions at the boundary~\Einstein.~%
\Footnote{Embacher
\Ref{Franz Embacher,
{\it Actions for signature change},
Phys.\ Rev.\ {\bf D51}, 6764 (1995).}
has derived field equations from a number of different versions of the
Einstein-Hilbert action, including the one given here.}

\Section{MATHEMATICS}

Theories involving internal boundaries are typically formulated using
distribution theory.  The standard theory of hypersurface distributions is
based on a nondegenerate volume element, which is usually taken to be the
metric volume element if available.  It is a remarkable property of
signature-changing spacetimes for which $\tau$ is an admissible coordinate
that, even though the metric is discontinuous, the (continuous extension of)
the metric volume element is smooth.  This of course follows immediately from
the smoothness of the generalized orthonormal frame, from which the volume
element can be constructed.  Thus, standard distribution theory can be used
with no further ado
\Ref{Tevian Dray, David Hartley, Robin W. Tucker, and Philip Tuckey, 
{\it Tensor Distributions in the Presence of Degenerate Metrics},
University of Adelaide preprint no.\ ADP 95-44/M37 (1995).}\label\Distribution
.%

Smooth signature-changing metrics, on the other hand, have metric volume
elements which vanish at $\Sigma$.  In fact, the combined requirements that
the metric volume element be used where possible and that smooth tensors be
distributions result in this case in a theory~\Distribution\ in which the
Dirac delta distribution is identically zero!

To illustrate these results, consider the following informal example.
Consider first the discontinuous signature-changing metric \Discontinuous\
with metric volume element
$$\omega = d\tau \wedge dx \Eqno$$
defined initially away from $\tau=0$, then continuously extended.  Let
$V=V^\tau\partial_\tau$ be a smooth vector field, and let
$${\bmit\delta} = d{\bmit\Theta} = \delta(\tau) \, d\tau \Eqno$$
be the standard hypersurface distribution associated with $\tau=0$, namely the
derivative of the Heaviside distribution ${\bmit\Theta}$.  Then
$$\langle{\bmit\delta},V\rangle
	= \int_M V^\tau \, \delta(\tau) \, \omega
	= \int_{t=0} V^\tau dx
  \Eqno$$
Now repeat the above construction for the smooth signature-changing metric
\Continuous\ with metric volume element
$$\hat\omega = \sqrt{|t|} \, dt \wedge dx \Eqno$$
again defined initially away from $t=0$, then continuously extended.  The
hypersurface distribution associated with $t=0$ is now
$${\bmit\delta} = d{\bmit\Theta} = \delta(t) \, dt \Eqno$$
so that if $\hat V=\hat V^t\partial_t$ is a smooth vector field then
$$\langle{\bmit\delta},\hat V\rangle
	= \int_M \hat V^t \, \delta(t) \, \hat\omega = 0 \Eqno$$ \label\Zero
since $\hat\omega=0$ at $t=0$.  The essential difference is not a change in
$\bmit\delta$, nor in the volume element, but rather fundamentally different
notions of what it means for the vector fields $V$ and $\hat V$ to be smooth.
For further details, see \Distribution.

This problem can of course be avoided for smooth signature-changing metrics by
using a nonmetric volume element.  For the above example, choosing the volume
element
$$\Omega = dt \wedge dx \Eqno$$
in the definition of distributions leads to \Zero\ being replaced by
$$\langle{\bmit\delta},\hat V\rangle
	= \int_M \hat V^t \, \delta(t) \, \Omega
	= \int_{t=0} V^t dx
  \Eqno$$
This theory is perfectly viable, and has been used to study the scalar field
on signature changing backgrounds.  However, the resulting distributions ---
foremost among them the Heaviside distribution --- differ from the
distributions one would naturally define on the Lorentzian region alone.
While this does not limit the usefulness of this approach, we find it
attractive that for discontinuous signature-changing metrics no such problem
arises.

\Section{DISCUSSION}

We have given both mathematical and physical examples of calculations which
are greatly simplified by working with generalized orthonormal frames when the
signature changes, and hence with proper ``time''~$\tau$.  Choosing a manifold
structure such that $\tau$ is a coordinate seems most likely to lead one
correctly through the minefield of choices one must make when dealing with a
degenerate metric.

Even in the constant signature case, while there is no need to use orthonormal
frames, many calculations become simpler if one does so.  One well-known
example is classical relativity itself, where the use of orthonormal tetrads
rather than, say, coordinate basis vectors, causes a vast reduction in the
number of independent components of the curvature tensor %
\Ref{See for instance:
\hfill\break
S. J. Campbell and J. Wainwright,
{\it Algebraic computing and the Newman-Penrose formalism
	in general relativity},
Gen.\ Rel.\ Grav.\ {\bf 8}, 987 (1977);
\hfill\break
Malcolm MacCallum,
{\it Comments on the performance of algebra systems in general relativity 
	and a recent paper by Nielsen and Pedersen},
ACM SIGSAM Bulletin, {\bf 23}, 22-25 (1989).}%
.  This fact formed the basis for the early work on the classification of
solutions of Einstein's equations using computer algebra; the coordinate-based
computations would have been too unwieldy.

The results described here for gravity are completely analogous to the work of
Dray {\it et al.}\ for the scalar field
\Ref{Tevian Dray, Corinne A. Manogue, and Robin W. Tucker,
{\it Particle Production from Signature Change},
Gen.\ Rel.\ Grav.\ {\bf 23}, 967 (1991);
\hfill\break
Tevian Dray, Corinne A. Manogue, and Robin W. Tucker,
{\it The Scalar Field Equation in the Presence of Signature Change},
Phys.\ Rev.\ {\bf D48}, 2587 (1993);
\hfill\break
Tevian Dray, Corinne A. Manogue, and Robin W. Tucker,
{\it Boundary Conditions for the Scalar Field
 in the Presence of Signature Change},
Class.\ Quantum Grav.\ {\bf 12}, 2767-2777 (1995).}%
, in which it was proposed that the field and its canonical momentum be
continuous at the surface of signature change.  Ellis and coworkers proposed
similar boundary conditions for both the scalar field and for gravity
\Ref{G Ellis, A Sumeruk, D Coule, C Hellaby,
{\it Change of Signature in Classical Relativity},
Class.\ Quant.\ Grav.\ {\bf 9}, 1535 (1992);
\hfill\break
G F R Ellis,
{\it Covariant Change of Signature in Classical Relativity},
Gen.\ Rel.\ Grav.\ {\bf 24}, 1047 (1992);
\hfill\break
Mauro Carfora and George Ellis,
{\it The Geometry of Classical Change of Signature},
Intl.\ J. Mod.\ Phys.\ {\bf D4}, 175 (1995).}%
.  Some of the implications of these boundary conditions for gravity
have been further explored by Hellaby and Dray
\Ref{Charles Hellaby and Tevian Dray,
{\it Failure of Standard Conservation Laws at a Classical Change of Signature},
Phys.\ Rev.\ {\bf D49}, 5096-5104 (1994);
\hfill\break
Tevian Dray and Charles Hellaby,
{\it The Patchwork Divergence Theorem},
J. Math.\ Phys.\ {\bf 35}, 5922-5929 (1994).}%
.%

\goodbreak
\bigskip
\leftline{\bf ACKNOWLEDGMENTS}
\nobreak
It is a pleasure to thank David Hartley, Marcus Kriele, J\"org Schray, Robin
Tucker, and Philip Tuckey for discussions.  This work was partially supported
by NSF Grant PHY-9208494 (CAM \& TD), two Fulbright Grants (TD, CAM), and a
research grant from the FRD (CH).

\vfill\eject
\References

\bye